\documentclass[conference]{IEEEtran}
\IEEEoverridecommandlockouts
\usepackage{cite}
\usepackage{amsmath,amssymb,amsfonts}
\usepackage{algorithmic}
\usepackage{graphicx}
\usepackage{textcomp}
\usepackage{xcolor}

\usepackage{listings}
\usepackage[utf8]{inputenc} % allow utf-8 input
\usepackage[T1]{fontenc}    % use 8-bit T1 fonts
\usepackage{url}            % simple URL typesetting
\usepackage{booktabs}       % professional-quality tables
\usepackage{amsfonts}       % blackboard math symbols
\usepackage{nicefrac}       % compact symbols for 1/2, etc.
\usepackage{microtype}      % microtypography
\usepackage{xcolor}         % colors

\usepackage{adjustbox}
\usepackage{floatrow}
\newfloatcommand{capbtabbox}{table}[][\FBwidth]
\usepackage{blindtext}

\usepackage{cite}
\usepackage{amsmath,amssymb,amsfonts}
\usepackage{graphicx}
\usepackage{textcomp}
\usepackage{xcolor}
\usepackage{comment}

\usepackage{subfig}
\usepackage{xcolor}
\usepackage{color,algorithm,algorithmic}
\usepackage{verbatim}
\usepackage{comment}
\usepackage{mwe}
\usepackage{url}
\usepackage{amsfonts,amsmath,amssymb,amsthm}
\usepackage{siunitx}
\usepackage{multirow,tabularx}

\theoremstyle{plain}

\theoremstyle{definition}

\theoremstyle{remark}

\providecommand{\delete}[1]{}
\providecommand{\nitro}[1]{}

\newcommand{\iid}{i.\@i.\@d.\@ }

\def\BibTeX{{\rm B\kern-.05em{\sc i\kern-.025em b}\kern-.08em
    T\kern-.1667em\lower.7ex\hbox{E}\kern-.125emX}}
    
\begin{document}

\title{Two-sample KS test with approxQuantile in \\
Apache Spark\textsuperscript{\textregistered}}

\author{\IEEEauthorblockN{Bradley Eck}
\IEEEauthorblockA{
\textit{IBM Research Europe}\\
Dublin, Ireland \\
bradley.eck@ie.ibm.com}
\and
\IEEEauthorblockN{Duygu Kabakci-Zorlu}
\IEEEauthorblockA{
\textit{IBM Research Europe}\\
Dublin, Ireland \\
duygu.kabakci.zorlu@ibm.com}
\and
\IEEEauthorblockN{Amadou Ba}
\IEEEauthorblockA{
\textit{IBM Research Europe}\\
Dublin, Ireland \\
amadouba@ie.ibm.com}
}

\maketitle

\begin{abstract}
The classical two-sample test of Kolmogorov-Smirnov (KS) is widely used to test whether empirical samples come from the same distribution. 
Even though most statistical packages provide an implementation,
carrying out the test in big data settings can be challenging because it requires a full sort of the data. 
The popular Apache Spark system for big data processing 
provides a 1-sample KS test, but not the 2-sample version.
Moreover, recent Spark versions provide the approxQuantile method for querying
$\epsilon$-approximate quantiles. 
We build on approxQuantile to propose a variation of the classical Kolmogorov-Smirnov two-sample test
that constructs approximate cumulative
distribution functions (CDF) from  $\epsilon$-approximate quantiles. 
We derive error bounds of the approximate CDF and show how to use this information
to carry out KS tests. Psuedocode for the approach requires 15 executable lines. 
A Python\textsuperscript{\textregistered} implementation appears in the appendix.
\end{abstract}

\section{Introduction}
The Kolmogorov-Smirnov (KS) test is widely used to test whether an empirical sample 
of data is consistent with a particular probability distribution \cite{PressTeukolsky88}.
The test works by computing the maximum distance between cumulative
distribution functions (CDFs). The one-sample version of the test compares the sample to
a theoretical distribution or model. The two-sample version compares the samples to 
each other and thus does not assume any particular underlying distribution. 

Two-sample KS tests have many applications. 
Our motivation relates to monitoring machine learning models to detect distribution
shifts among model features and targets, especially in big data settings \cite{eck2022}. 
Machine learning models usually assume stationarity of distributions between data
used in training and that encountered for production \cite{Ditzler15}. 
A two-sample test can help identify violations of the stationarity assumption. 
Moreover, the supply chain forecasting application that motivated our study
makes extensive use of Apache Spark for data processing and machine learning.

Apache Spark library finds considerable use for computing with big data and contains many useful routines applicable to model monitoring.
These include a 1-sample KS test (\texttt{KolmogorovSmirnovTest}), which compares a sample of observations to a normal distribution.
However, the two-sample test is not supported as of pyspark 3.4.1 (June 2023).
Users wishing to carry out a two-sample test could implement it from scratch. 
However, implementing the test from scratch requires storing a full sort of the data and this is not always feasible or desired. 
As an alternative, we propose the use of Spark's \texttt{approxQuantile} method to build approximate CDFs which can then be used for KS testing. 

The contributions of our paper are the following. \textbf{(1)} We build approximate cumulative distribution functions (CDF) from $\epsilon$-approximate quantiles. \textbf{(2)} We derive the error bound of the approximate CDF and provide a heuristic for parameterizing the CDF to obtain 
a desired error bound.
\textbf{(3)} We conduct Kolmogorov-Smirnov (KS) tests to detect distribution shifts using the approximated CDF and error bound.
\textbf{(4)} We provide a psuedocode and Python code for the routine comprising only a few executable lines.

The paper is organized as follows: section \ref{sec:background} summarizes 
background information and related work. 
Our approach to determining an approximate CDFs and carrying out KS tests is exposed in section \ref{methodology}.
Section \ref{experiments} validates our approach with experiments. Finally, section \ref{conclusions} summarizes the concluding remarks.

\section{Background}
\label{sec:background}

\subsection{Two-sample Kolmogorov-Smirnov Test}

The KS test provides a means for testing the hypothesis that two
samples are drawn from the same distribution. The test statistic $D_{KS}$ is the maximum 
distance between the CDF of each sample.
\begin{equation}
D_{KS} = | F_1(x) - F_2(x) |_\infty
\label{eq:DKS}
\end{equation}
In big data settings, evaluating Equation \ref{eq:DKS} can be challenging 
as it requires storing a full sort of the data.
The test statistic has the useful properties of having a well known 
distribution.
The significance level of an observed $D_{KS}$ distance is approximated
by \cite{PressTeukolsky88}: 

\begin{equation}
  P(D_{KS}>observed) = Q_{KS}\bigg(\sqrt{\frac{N M}{N+M}}D_{KS}\bigg)
  \label{eq:PDks}
\end{equation}
where the quantile function is:
\begin{equation}
  Q_{KS}(\lambda) = 2 \sum_{k=1}^{\infty} (-1)^{k-1} e^{-2k^2 \lambda^2}
\end{equation}

\subsection{KS Variations}
In the area of industrial deployments of machine learning models that motivates our work, several recent papers use KS tests to study distribution shift or concept drift.
As an example, \cite{Rabanser2019} showed how KS test can be used for detecting distribution shift, and that KS tests remain competitive even in higher dimensions.
The work of \cite{Glazer} proposed a generalized, nonparametric, statistical KS test for detecting changes in PDF and in high-dimensional data.
\cite{Cong} proposed an intuitive way to arrive at explanations in KS test which accommodates both the KS test data and the user domain knowledge in producing explanations.
Along the same line, \cite{does} developed an incremental KS algorithm that allows performing the KS hypothesis test instantly using two samples that change over time, where the change is an insertion and/or removal of an observation.
In a contribution similar to the present work, \cite{Lall} applied the data structure of \cite{Greenwald2001} to carry out KS tests on streaming data. 
Despite the long history, recent citations indicate the continued popularity of applying KS tests in new situations. Our approach falls in this category.

\subsection{$\epsilon$-approximate Quantiles}

A closely related problem to finding cumulative distribution functions needed
by Eq.(\ref{eq:DKS}) is that of making quantile estimates.
For data size $N$,  a quantile estimate is termed \textit{$\epsilon$-approximate} when it has precision of $\epsilon N$. 
Thus for
any requested rank $\hat{r}$, an $\epsilon$-approximate quantile summary returns a value
from the data sequence whose true
rank $r$ is guaranteed to be within the interval $\left[r-\epsilon N, \, r+\epsilon
N\right]$.
A variant of this relation allows to query quantiles. For example, if we request the
quantile at probability $\hat{p}$ up to an error $\epsilon$, then the summary will
return a sample $x$ from the sequence of $N$ so that the exact rank of $x$ is close to
$\hat{p} \times N$ and follows the relation\cite{approxQuantile}:
% \cite{spark}
\begin{equation}
    floor((\hat{p} - \epsilon) * N) \leq rank(x) \leq ceil((\hat{p} + \epsilon) * N)
\label{eq:approxQuantile}
\end{equation}

Construction of a sketch or data summary that can answer such quantile queries 
is non-trivial. 
The work of  \cite{ZhangWang2007} proposed an algorithm to approximate quantiles in
the context of streaming data. And, \cite{guha} developed a method for estimating quantiles in random order statistics. 
The now classical algorithm of
\cite{Greenwald2001} provides a state of the art approach with the lowest memory requirement $O(\frac{1}{\epsilon} log (\epsilon N))$.  
This method also forms the basis of the approxQuantile method provided by Spark \cite{Spark}.

\section{Methodology}
\label{methodology}

We present in this section our approximate CDF using quantile estimates. We then analyze the approximation error and propose a heuristic to select a trade-off between the discretization of the CDF and the accuracy of the quantile estimates. The analysis leads to a routine for KS test.

\subsection{Approximate CDFs}

\begin{figure}
\centering
\includegraphics[width=0.8\textwidth]{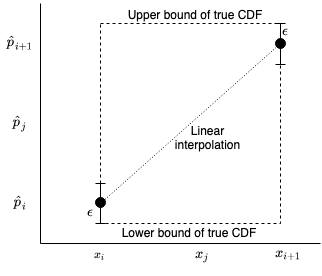}
\caption{Bounds on true CDF between $\epsilon$-approximate quantiles.}
\label{fig:approxQuantile}
\end{figure}

In this work, we build upon the $\epsilon$-approximate quantiles from \cite{Greenwald2001} as exposed in Spark's approxQuantile method to create
an approximate CDF. 
We construct an approximate CDF,  $\hat{F}(x)$, of the true cumulative distribution $F(x)$ using
quantiles $x_i$ queried from the quantile summary at probabilities
$1/N \leq \hat{p}_i \leq 1$.
The pairs of approximated quantiles and requested probabilities,  $(x_i, \hat{p}_i)$ can be used to evaluate
 $\hat{F}(x)$ by interpolating between the points. Since $F$ is always increasing,
 any monotonic interpolating function will  work. We choose a linear interpolation between the points
 for simplicity. An approximation for the true CDF is given by:

\begin{equation}
\hat{F}(x_j) = \hat{p}_i + \frac{x_j - x_i}{x_{i+1}-x_i} (\hat{p}_{i+1} - \hat{p}_i); \,  x_i \leq x_j \leq x_{i+1}
\label{eq:fhat}
\end{equation}
where the subscript $i$ runs over the pairs of approximate quantiles
and requested probabilities and the $x_j$ denotes an observation for which we
evaluate $\hat{F}(x)$. 
The error in $\hat{F}$ depends on both the relative error applied to the quantile
estimates, $\epsilon$, and the spacing between probabilities at which the quantiles
were estimated, $\Delta P = \hat{p}_{i+1} - \hat{p}_i$.  The error in the probability $\hat{p}_i$ at sample
quantile $x_i$ is $\epsilon$. Between these approximate quantiles, the true
probability lies within a rectangle with lower left corner 
$(x_i, \hat{p}_i - \epsilon)$ 
and upper right corner 
$(x_{i+1}, \hat{p}_{i+1} + \epsilon)$ (Figure \ref{fig:approxQuantile}).

As the actual value of $F(x)$ lies in this rectangle, the error in $\hat{F}(x)$ can be bounded as the difference between the estimate $\hat{F}(x)$ and the true $F(x)$. Considering $F(x) > \hat{F}(x)$ and denoting the error bound as $\delta$, we can write $\delta \le F(x) - \hat{F}(x) = \hat{p}_{i+1}+\epsilon - \hat{F}(x)$. Using (\ref{eq:fhat}) and writing $\Delta P=\hat{p}_{i+1} -\hat{p}_i$, we obtain $\delta \le \Delta P (1 - \frac{x_j - x_i}{x_{i+1}-x_i}) + \epsilon$. The fact that the parenthetical term will be at most 1, and $3 \le a \le N$ equi-spaced points leads to $\Delta P = \frac{1}{a-1}$, we can write:

\begin{equation}
  \delta \leq \frac{1}{a-1} + \epsilon
\label{eq:delta}
\end{equation}
By symmetry, the error bound is the same for $F(x) < \hat{F}(x).$ 
Eq. (\ref{eq:delta}) can be used
to select $a$ and $\epsilon$, for a given value of $\delta$.
For a chosen $\delta$, we may choose between
more accurate quantile estimates and more points.
This trade-off between the number of points $a$ and the value of $\epsilon$ follow
the hyperbola $a=\frac{1}{\delta-\epsilon}+1$. 

We propose the following a heuristic to enable closed form selection of $a$ and
$\epsilon$ for a given $\delta$.
First, we make a change of variables to $y=a/N$ and
$x=\epsilon / \delta$.  In these coordinates, the curve has a slope near $0$
at low values of $\epsilon$, moving toward infinity (90 degrees) as $\epsilon$ approaches
the asymptote at $\epsilon = \delta$ or $x=1$.  We choose
the point where the slope is unity, 45 degrees:

\begin{equation}
\epsilon_{45} = \delta - \sqrt{\frac{\delta}{N}}
\label{eq:eps45}
\end{equation}

\begin{figure}
  \centering
  \includegraphics[width=\textwidth]{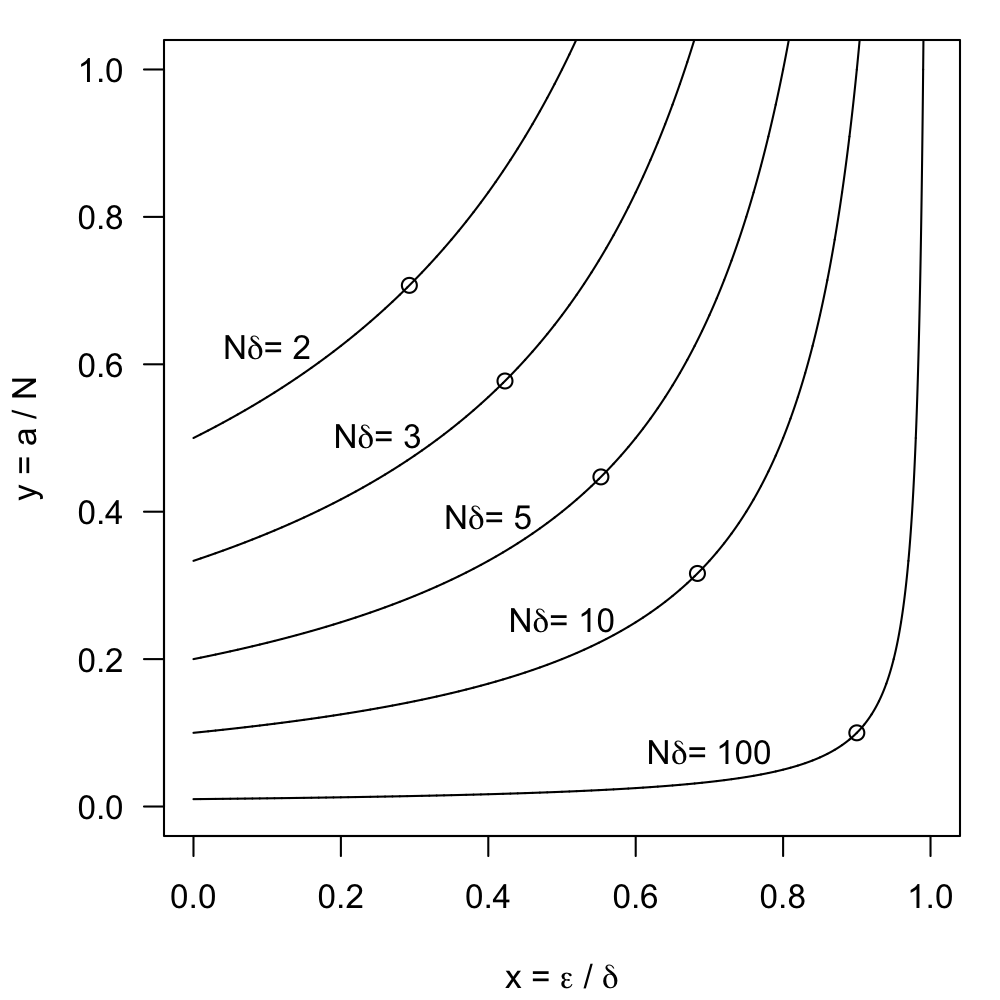}
  \caption{Tradeoff between number of points for approximating CDF and
          the error in quantile estimates. Circles show point of unit slope.}
  \label{fig:tradeoff}
\end{figure}

Figure \ref{fig:tradeoff} shows hyperbolas of the trade-off between accuracy 
of quantile queries and the number of points used in the CDF.

\subsection{Approximate 2 sample KS}

To carry out KS testing with approximate CDFs, we compute 
the test statistic (\ref{eq:DKS}) using the approximate CDFs developed above.

The approximate CDFs comprise the vector of requested probabilites and their
approximate quantiles. Linear interpolation is used to evalute the CDF to find
cumulative probabilities for other values.

To compute the KS distance, we evaluate the reference CDF, $\hat{F}_1$, at
quantiles $y_j$ stored for the test CDF and evaluate the test CDF, $\hat{F}_2$,
at quantiles $x_i$ stored for the reference CDF.  The largest absolute
difference between probabilties is our estimate of the KS distance:
 
\begin{equation}
\hat{D}_{KS} = | \hat{F}_1(x_i) - \hat{F}_2(x_i) \cup (\hat{F}_1(y_j) - \hat{F}_2(y_i)|_{\infty}
\end{equation}

The error in $\hat{D}_{KS}$ is bounded by the sum of the error bounds for 
$\hat{F}_1$ and $\hat{F}_2$: $\delta_1 + \delta_2$.

Applying CDF approximations to KS testing requires selection of the approximate CDF 
parameters from the test conditions. 
Here we follow the suggestion of \cite{Lall} and choose $\phi$ so that 
the error introduced by using the approximate CDFs is 
"small enough not to degrade the quality of the test". 
One way of specifying the quality of the test
is defining the precision, $\beta$ of the resulting p-values along with the significance level.
For KS testing, we choose $\phi$ as the smaller of the absolute differences
in critical distance $D_{crit}$ between the target $\alpha$ and $\alpha \pm \beta$: 
\begin{equation}
   \phi  = min |D_{crit}(\alpha \pm \beta, N,M) - D_{crit}(\alpha, N,M)|
   \label{eq:phi}
\end{equation}
where $D_{crit}$ is obtained by inverting \ref{eq:PDks} for $\alpha$.
Using $\phi$ and $N$, $\epsilon$ can be obtained from (\ref{eq:eps45})
and $a$ from (\ref{eq:delta}).

Psuedo code for our approximate 2-sample KS test appears in figure
\ref{fig:algo}.  The routine takes as input a required precision in $D_{KS}$ of
$\phi$ and sample vectors $x$ and $y$, with size $N$ and $M$. 
First, find the parameters of the approximate CDF for each sample. 
At line 1, the error of each CDF, $\delta$,  should be less than $\frac{\phi}{2}.$
Lines 2-3 find a suitable value of $\epsilon$ for approximating quantiles 
using Equation \ref{eq:eps45}.
Lines 4-7 create arrays of linearly spaced probabilities for each sample.
Approximate quantiles of the samples are computed in lines 8-9.
Evaluation of the approximate CDFs using Equation \ref{eq:fhat} appears in 10-13. 
Finally the KS distance is calculated at line 14 and returned.

\begin{figure}[]
\begin{flushleft} 
\textbf{approx2sampKS}($\phi, x, N, y, M$ )
\end{flushleft} 
\begin{algorithmic}[1]
\STATE $\delta \gets \phi/2$
\STATE $\epsilon_{45x} \gets \delta - \sqrt{\frac{\delta}{N}}$
\STATE $\epsilon_{45y} \gets \delta - \sqrt{\frac{\delta}{M}}$
\STATE $a_x \gets \lceil \frac{1}{\delta-\epsilon_{45x}} + 1 \rceil$
\STATE $a_y \gets \lceil \frac{1}{\delta-\epsilon_{45y}} + 1 \rceil$
\STATE $\hat{px}_i \gets linspace(1/N,1,a_x)$
\STATE $\hat{py}_j \gets linspace(1/M,1,a_y)$
\STATE $x_i \gets approxQuantile(x, \hat{px}_i, \epsilon_{45x})$
\STATE $y_j \gets approxQuantile(y, \hat{py}_j, \epsilon_{45y})$
\STATE $\hat{Fx}_i \gets \hat{px}_i $
\STATE $\hat{Fy}_i \gets interp(x_i, y_j, \hat{py}_j)$
\STATE $\hat{Fy}_j \gets \hat{py}_j$
\STATE $\hat{Fx}_j \gets interp(x_i, y_j, \hat{py}_j)$
\STATE $\hat{D}_{KS} \gets | (\hat{Fx}_i - \hat{Fy}_i) \cup (\hat{Fx}_j - \hat{Fy}{j}) |_{\infty} $
\begin{flushleft} 
\textbf{return} $\hat{D}_{KS}$
\end{flushleft} 
\end{algorithmic}
\caption{Psuedocode code for two-sample KS test with Spark's approxQuantile. $\phi$ is 
the required precision in $D_{KS}$. Samples $x$ and $y$ have sizes $N$ and $M$.}
\label{fig:algo}
\end{figure}

\section{Experiments}
\label{experiments}
This section reports computational experiments that validate the approximate CDF 
and KS test presented above. Experiments were carried out in Python 3.9 on a
Macbook Pro 2.6 GHz 6-Core Intel Core i7 with 16 GB of memory. Library code
used in the experiments included pyspark, numpy, scipy, and matplotlib.

\subsection{Convergence study for CDF approximation error}

Before applying the approximate CDF  
for KS tests, we first study the numerical performance of the approximation.
We generated 20 samples of 10,000 points each from the standard normal distribution.
Approximate CDFs were computed for each sample using $\epsilon={0.1, 0.01, 0.001}$
and three values of $a$. We evaluated the approximate CDFs at all 10,000
points and the reported probability compared to the exact value obtained from a 
full sort of the sample. The largest magnitude error, max|error|, over the 20 samples
appears in each row of Table \ref{tab:convergence}. All of the observed errors 
fall within the theoretical bound of $\delta$ from (\ref{eq:delta}).

\begin{figure}[h!]
    \centering
    \includegraphics[width=\textwidth]{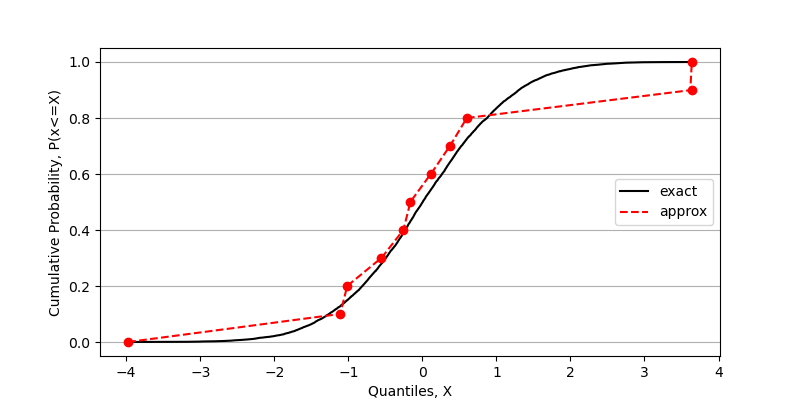}
    \caption{Comparison of exact and approximate CDF for a convergence experiment. $\epsilon=0.1$ and $a=11$.}
    \label{fig:cdf}
\end{figure}

Plots of the exact and approximate CDF from one of the convergence experiments
(Figure \ref{fig:cdf}) illustrate the approach. The exact curve contains 10,000 
sample points. The red dots show $\epsilon$-approximate quantiles returned 
at 11 linearly spaced points between the sample minimum and maximum. Most 
of the approximate points fall near the curve but one point, second from right,
stands out. Although this point falls further from the exact curve it is 
nonetheless within the 
specified tolerance of 0.1 from the requested probability 0.9. Considering this point
for linear interpolation of the CDF according to (\ref{eq:fhat}) shows an experimental
example of the theoretical error bound of Figure \ref{fig:tradeoff}. 

\begin{table}[]
    \centering
    \caption{Convergence study for CDF approximation error; max|error| computed
             over 20 experiments each with 10,000 points generated from the 
             standard normal distribution. All observed errors fall within
             error bound $\delta$ computed from the approximation parameters
             $a$ and $\epsilon$.}
    
    \begin{tabular}{rlll}
    
    \hline
%     \multicolumn{2}{c}{Approximation Parameters} & Error Bound & Observed Error \\
      $a$ & $\epsilon$ & $\delta$  &  max |error| \\
      \hline
      11  &  0.1       & 0.20      &   0.1456  \\
      21  &  0.1       & 0.15      &   0.1208  \\
      51  &  0.1       & 0.12      &   0.1069  \\
      11  &  0.01      & 0.11      &   0.05692  \\
      21  &  0.01      & 0.06      &   0.02867  \\
      51  &  0.01      & 0.03      &   0.01186  \\
      101 &  0.001     & 0.011     &   0.005782  \\
      201 &  0.001     & 0.006     &   0.002725  \\
      501 &  0.001     & 0.003     &   0.001714  \\
%    N(0,1)        & 1e5        & 101 &  0.001     & 0.011     &   0.006304 \\
%      N(0,1)        & 1e5        & 201 &  0.001     & 0.006     &   0.003093  \\
%      N(0,1)        & 1e5        & 501 &  0.001     & 0.003     &   0.001260  \\
    \hline
    \end{tabular}
 
    \label{tab:convergence}
\end{table}

\subsection{KS Testing with approximated CDF}

\begin{table*}[]
    \centering
    \caption{Summary results for Kolmogorov-Smirnov experiments: range of exact
             distances, $D_{KS}$, range of approx distances $\hat{D}_{KS}$,
             maximum absolute distance error $|\hat{D}_{KS} - D_{KS}|_{\infty}$, p-values of $D_{KS}$ and $\hat{D}_{KS}$ as $[P (D > D_{KS})]$ and $[P( D > \hat{D}_{KS})]$, number of $H_{0}$ rejections of $D_{KS}$ and $\hat{D}_{KS}$
            for five experiments each with 20 replications. Experiments 1-3 had equal 
            sample sizes of $10^4$. Experiments 4 and 5 had reference samples of 84,000
            and test samples of 7,000.}
    \begin{adjustbox}{width=\linewidth,center}
    \begin{tabular}{llllll}
    \hline
     Experiment      & 1      &  2     &   3    &     4           &     5           \\         
     Sample 1     & $\mathcal{N}$(0,1) & $\mathcal{N}$(0,1) & $\mathcal{N}$(0,1) & $\Gamma(0.5,1)$ & $\Gamma(0.5,1)$ \\
     Sample 2     & $\mathcal{N}$(1,1) & $\mathcal{N}$(0,2) & $\mathcal{N}$(0,1) & $\mathcal{U}$(0,1)          & $\Gamma(0.5,1)$ \\
    \hline
$[D_{KS}]$       & 0.3684, 0.4007 & 0.0837, 0.0976 & 0.0072, 0.0220 & 0.2657, 0.2850 & 0.00624, 0.01874 \\
$[\hat{D}_{KS}]$ & 0.3684, 0.4006 & 0.0837, 0.0976 & 0.0071, 0.0219 & 0.2656, 0.2849 & 0.00625, 0.01859 \\
$|\hat{D}_{KS} - D_{KS}|_{\infty}$ & 0.000162 & 0.000159 & 0.000147 & 0.000120 & 0.000217 \\
$[P (D > D_{KS})]$ & 0.0, 0.0       & 7.4e-42, 7.0e-31 & 0.01581, 0.9578 & 0.0, 0.0 & 0.02112, 0.9616 \\
$[P( D > \hat{D}_{KS})]$ & 0.0, 0.0 & 9.2e-42, 7.1e-31 & 0.01656, 0.9628 & 0.0, 0.0 & 0.02297, 0.9626\\
$ \# (P( D > {D}_{KS}) <= \alpha ) $ & 20 & 20 & 1 & 20 & 8\\
$ \# (P( D > \hat{D}_{KS}) <= \alpha)$ & 20 & 20 & 1 & 20 & 7\\
\hline
    \end{tabular}
    \end{adjustbox}
    \label{tab:KS}
\end{table*}

We report experiments for two samples KS testing using our 
approximate CDF in two settings: Gaussian samples of the same 
size and Gamma and Uniformly distributed samples of different sizes.

For the Gaussians, we perform a two-sample KS test at significance level
$\alpha=0.05$, where the reference and test samples have $10^4$ points. 
Considering a precision of 0.025, (\ref{eq:phi}) gives a value of $\phi= 0.000399$.
We construct approximate CDF using this error bound where
$\epsilon=5.8e-5$ from (\ref{eq:eps45}) and $a=7083$ from (\ref{eq:delta}).
Under these conditions we carried out three experiments of 20 replications
each.  First, we tested for a shift in mean using a reference sample $~
\mathcal{N}\left(0, 1\right)$ and test sample $~ \mathcal{N}\left(1, 1\right)$.
Second, we tested for shift in variance with reference sample $~
\mathcal{N}\left(0, 1 \right)$ and test sample $~ \mathcal{N}\left(0,
2\right)$. Third, we tested the case where no shift was present with reference
sample $~ \mathcal{N}\left(0, 1 \right)$ and test sample $~ \mathcal{N}\left(0,
1 \right)$. 

A second set of experiments using Gamma and Uniform distributions having different
sample sizes is motivated by the modeling problem of forecasting inventory demand
in retail settings.
A deep learning model predicts
the demand for thousands of products using tens of features.  The model is
(re)trained monthly using the previous 3 months of data. With around 1000
observations per day, the training data is around $N=12*7*1000=84,000$ (per
product). Monitoring for distribution shift runs weekly $N=7000$ in order to detect a distribution
shift at a significance level $\alpha=0.20$ with precision $0.1$.
Approximate CDF were constructed using an error bound $\delta = \frac{\phi}{2}=0.000385$ from (\ref{eq:phi}).
For the reference sample (N=84,000) we have $\epsilon= 0.00032$, and $a=14770$ 
from Eqs. (\ref{eq:eps45}) and (\ref{eq:delta}), respectively.
Similarly for the test sample (N=7000) we have $\epsilon=0.00015$ and $a=4265$. Under these
conditions we again carried out 20 replications each.  The fourth KS experiment
considers a reference distribution drawn from the Gamma family
(shape=0.5, scale=1) and a target distribution drawn from uniform (min=0, max=1).
The fifth and KS experiment reported here compares samples of different sizes
drawn from the same gamma distribution with scale 0.5 and shape 1.0.

Table \ref{tab:KS} summarizes the test settings and results. Across all experiments
the approximate distance compares favorably with the exact distance. Furthermore, 
the worst distance error always fell within the expected bound for a two sample 
test of $2\delta$. 

We evaluate p-values of exact KS test and approximate KS test to accept or
reject null hypothesis, two samples were drawn from the same distribution, by
comparing significance level. If p-value of test is less than significance
level, null hypothesis can be rejected. All 20 repetitions of first, second and
fourth setups result in rejection of null hypothesis with 0.05 and 0.20
significance levels respectively. Only  fifth experiment has single
disagreement between approximate and exact KS tests that fell into expected
error level due to density function approximation of 0.05 precision.

We also conduct experiments on real data obtained from
Columbia University’s EnHANTs (Energy Harvesting Active Networked Tags)
project, \cite{columbia-enhants-20110407}. We compared irradiance measurements
(in units $W/cm^2$) of Traces A and Traces B sizes 1,107,796 and 984,000
respectively after cleaning 25,840 and 97,792 missing data
points from original traces. To perform two sample approximate KS Test,
approximate CDFs were constructed using error bound $\delta=0.001$.
For the reference sample Trace A,  we have $\epsilon= 0.000970$, and $a=33,285$
from (\ref{eq:eps45}) and (\ref{eq:delta}), respectively. 
Similarly for the test sample, we have $\epsilon=0.000968$ and
$a=31,370$. We obtain similar approximate KS with exact KS results which are
$D_{KS}=0.22945$ and $\hat{D}_{KS}=0.26486$ respectively. P-values of tests are
calculated as 0.0 which states clear rejection of null hypothesis.

\subsection{Comparison with Lall's KS Test}

\begin{table*}
\caption{Comparison of KS tests for five synthetic experiments.}
\begin{tabular}{lrrrrr}
\hline
Experiment                 & 6 & 7& 8 & 9 & 10  \\
Sample 1  & $\mathcal{N}$(0,1) & $\mathcal{N}$(0,1) & $\mathcal{N}$(0,1) & $\Gamma(0.5,1)$ & $\Gamma(0.5,1)$ \\
Sample 2  & $\mathcal{N}$(1,1) & $\mathcal{N}$(0,2) & $\mathcal{N}$(0,1) & $\mathcal{U}$(0,1)          & $\Gamma(0.5,1)$ \\
\hline
Sample size                & 10,000 & 10,000 & 100,000 & 84,000 & 84,000 \\
Target Precision in $D_{KS}$ & 0.05 & 0.01 & 0.001 & 0.05 & 0.002 \\
$\delta$ for CDF           & 0.025& 0.005 & 0.0005 & 0.025 & 0.001\\ 
$\epsilon_{45}$ for CDF    & 0.02341&0.004293& 0.000429 & 0.02445 & 0.000892\\
$\epsilon$ for Sketch      & 0.00833&0.001667 & 0.000167 &0.00833 & 0.000333\\
\hline
CDF size                   & 634    & 1416 & 14144 & 1835 & 9167 \\
Sketch size                 & 131  & 607 &  6000 & 157 & 3949  \\
%Sketch "bound"            & 230 & 733 & 1331 & 230 & 1568 \\
\hline
$D_{KS}$ (exact) & 0.3684 & 0.0976 & 0.00426 & 0.2751 & 0.00432 \\
$D_{KS}$ via CDF  & 0.3786 & 0.0992 & 0.00428 & 0.2906 & 0.00440 \\ 
$D_{KS}$ via sketch \cite{Lall} & 0.3857 & 0.1003 & 0.00460 & 0.2850 & 0.00488 \\
\hline
\end{tabular}
\label{tab:lall}
\end{table*}

We compare our KS test using an approximate CDF with the closely related method proposed by Lall \cite{Lall}.
That method uses the $\epsilon$-sketch of \cite{Greenwald2001} directly to carry out the KS test. Whereas our method uses the sketch as an intermediate step, via Spark's approxQuantile method. To facilitate the comparison, we implemented the $\epsilon$-sketch of \cite{Greenwald2001} in order to obtain direct access to the sketch itself. 

We compare the tests using the same precision in $D_{KS}$ (Table \ref{tab:lall}). 
For the approximate CDFs, we choose $\delta$ as
half of the chosen precision in $D_{KS}$ based on the bound developed in section \ref{methodology}. The value of $\epsilon_{45}$ comes from equation \ref{eq:eps45}.
The value of $\epsilon$ is chosen as $\frac{1}{6}$ of the precision in $D_{KS}$--the bound reported by \cite{Lall}.
Values of the test statistic, $D_{KS}$,
obtained by both methods are near the exact value, and within the target precision specified for the experiments. 

It is interesting to compare the storage requirements between the methods.
The approximate CDFs use a larger value of $\epsilon$ than Lall's method.
Based on this difference, one might expect the sketches to need more points.
However, our results show that the CDFs require more points than the sketches.
This result matches the expected scaling of space complexity.
The approximate CDFs scale as $O(\sqrt{N})$ whereas the $\epsilon$ sketch of \cite{Greenwald2001} scales as $O(\frac{1}{\epsilon} log( \epsilon N)$. 
The larger values of $\epsilon$ used by the CDFs are not enough larger to overcome
the log scaling of the sketch.

\section{Conclusions}
\label{conclusions}
This paper developed an approximated CDF from the existing work of an $\epsilon$-approximate quantile summary, derived the error bound for the approximated CDF, and confirmed this bound by experiment. 
We applied the CDF and its error bound to the classical technique of KS testing, obtaining the same results as carrying out a full KS test, but using fewer points thanks to the approximation. Furthermore, the number of points needed can be tuned based on the desired significance level and precision of the KS test. 

Our experiments also compared with a closely related existing approach that uses a sketch directly.
Given the superior space complexity of using the sketch directly, and the fact
that approximate CDFs anyhow use a sketch as an intermediate step, one might
ask why approximate CDFs should be used for KS testing.
Our reply would be that the straightforward implementation using the approxQuantile 
makes it easy to bring two-sample KS testing to existing environments using Spark.
In such cases the sketch may not be accessible through the available APIs
and computing a new sketch would require an implementation thereof, which is not trivial,
and processing of the full data, which may be undesireable.
In such cases the method proposed here can be used to carry out KS tests with 
good accuracy.

\bibliographystyle{IEEEtran}
\newpage
\bibliography{approxKS}

\appendix[Python code for approx2sampKS]
This code is available under the MIT License \cite{MIT}.
\begin{scriptsize}
\begin{verbatim}
from pyspark.sql import DataFrame
from math import ceil, sqrt
from numpy import linspace, interp

def approx2sampKS(phi:float, 
                  xdf:DataFrame, xcol:str, 
                  ydf:DataFrame, ycol:str) -> float:
    """Approximates two-sample KS distance with precision
       phi between columns of Spark DataFrames."""
    delta = phi / 2
    N = xdf.count()
    eps45x = eps45(delta,N)
    M = ydf.count()
    eps45y = eps45(delta,M)
    ax = num_probs(N, delta, eps45x)
    ay = num_probs(M, delta, eps45y)
    pxi = linspace(1/N,1,ax)
    pyj = linspace(1/M,1,ay)
    xi  = xdf.approxQuantile(xcol, list(pxi), eps45x)
    yj =  ydf.approxQuantile(ycol, list(pyj), eps45y)
    Fxi = pxi
    Fyi = interp(xi, yj,pyj)
    Fyj = pyj
    Fxj = interp(yj, xi, pxi)
    Di = max(abs( Fxi - Fyi))
    Dj = max(abs( Fxj - Fyj))
    D_KS = max( Di, Dj)
    return D_KS

def eps45(delta, N):
    """Select value of epsilon at the elbow of unit slope.
       Return 0, for no approximation if delta < 0"""
    eps = max(0.0, delta - sqrt(delta/N)) 
    return eps

def num_probs(N:int, delta:float, epsilon:float):
    """Calculate number of probability points for approx
       quantiles; at most this is the number of data points."""
    a = 1 / (delta - epsilon) + 1
    return min(ceil(a), N)
\end{verbatim}
\end{scriptsize}

\end{document}